\documentclass[aps,prd,a4paper,reprint,amsmath,amssymb,superscriptaddress,nofootinbib]{revtex4-2}

\usepackage{graphicx}
\usepackage{dcolumn}
\usepackage{footmisc}
\usepackage[normalem]{ulem}
\usepackage{bm}
\usepackage{hyperref}
\usepackage{xcolor}
\usepackage{float}
\usepackage{hyperref}

\begin{document}
\setlength{\abovedisplayskip}{5pt}
\setlength{\belowdisplayskip}{5pt}
\setlength{\abovedisplayshortskip}{5pt}
\setlength{\belowdisplayshortskip}{5pt}
\hyphenpenalty=1050

\preprint{}

\title{Polarization Measurements as a Probe of Axion-Photon Coupling:\\ a Study of GRB 221009A}

\author{Boris Betancourt Kamenetskaia}
\affiliation{Technical University of Munich, TUM School of Natural Sciences, Physics Department, Professorship T30d Elementary Particle Physics, James-Franck-Strasse 1, 85748 Garching, Germany,}
\affiliation{Max-Planck-Institut f\"{u}r Physik (Werner-Heisenberg-Institut), Boltzmannstra\ss e 8, 
85748 Garching, Germany,}
\email{boris.betancourt@tum.de}

\author{Nissim Fraija}
\affiliation{Instituto de Astronom\' ia, Universidad Nacional Aut\'onoma de M\'exico, Circuito Exterior, C.U., A. Postal 70-264, 04510 Cd. de M\'exico,  M\'exico}
\email{nifraija@astro.unam.mx }

\author{Gonzalo Herrera}
\affiliation{Center for Neutrino Physics, Department of Physics,\\ Virginia Tech, Blacksburg, Virginia 24061, USA}
\email{gonzaloherrera@vt.edu}

\begin{abstract}
Axionlike Particles (ALPs) can be produced in gamma ray bursts, altering the polarization of the electromagnetic emission in these events. For the first time, we derive bounds on the axion-photon coupling from polarization measurements of GRB 221009A, performing a full calculation of the Stokes parameters, as it is typically done in the astrophysics community. Within astrophysical uncertainties, our limits on the axion-photon coupling are competitive with complementary probes in the axion mass range $10^{-9}$ eV $\lesssim m_a \lesssim 10^{-8}$ eV, further allowing to probe motivated parameter space of ALP dark matter. 

\end{abstract}

\maketitle

\section{INTRODUCTION}\label{sec:intro}

Pseudo Nambu-Goldstone bosons of a global U(1) symmetry, called ``axions," constitute a solution to the strong \textit{CP} problem \cite{Peccei:1977hh, Peccei:1977ur,Weinberg:1977ma, Wilczek:1977pj}. Generalized ``axionlike particles" (ALPs) with similar features to axions but unable to solve the strong \textit{CP} problem appear in multiple extensions of the Standard Model, and can account for the dark matter of the Universe under certain conditions \cite{Abbott:1982af,Dine:1982ah,Preskill:1982cy,Turner:1986tb, Sikivie:2006ni}. Axions can mix with photons in the presence of an electromagnetic field, and axion-photon conversion can occur \cite{Raffelt:1985nj, PhysRevD.37.1237, Mirizzi:2007hr}.
Astrophysical searches for axions often focus on such conversions, making stars, supernovae, and galaxy clusters natural laboratories for these investigations. For instance, the Sun is a prominent target in axion searches, as axions produced in the solar core via the Primakoff effect \cite{Primakoff:1951iae,PhysRevD.37.1237} might convert into x-ray photons within the magnetic fields surrounding detectors on Earth \cite{CAST:2017uph}. Supernovae, particularly SN 1987A, have also been studied for axion emissions, where the cooling rates of neutron stars provide constraints on the axion mass \cite{Raffelt:1987yt,SN1987AII}. Additionally, galaxy clusters are prime candidates for axion searches due to their large-scale magnetic fields, where axion-photon conversion could be detected as x-ray excesses or anomalies \cite{Wouters:2013hua}, offering a unique observational window into axion properties.

The starburst galaxy M82 is also of interest in these searches due to its intense star formation and associated magnetic activity. Observations targeting x-ray or gamma-ray emissions from M82 can therefore provide constraints on ALP properties \cite{Ning:2024eky}. Similarly, neutron stars, especially magnetars with their extremely strong magnetic fields (up to $10^{15}~\rm G$), offer another promising environment for axion searches. Axions produced in the dense core of a neutron star could convert into photons in the star's magnetosphere, leading to observable x-ray or radio signals \cite{Pshirkov:2007st,Buschmann:2019pfp,Song:2024rru}.

\begin{figure}[H]
	\centering
\includegraphics[width=1.0\linewidth]{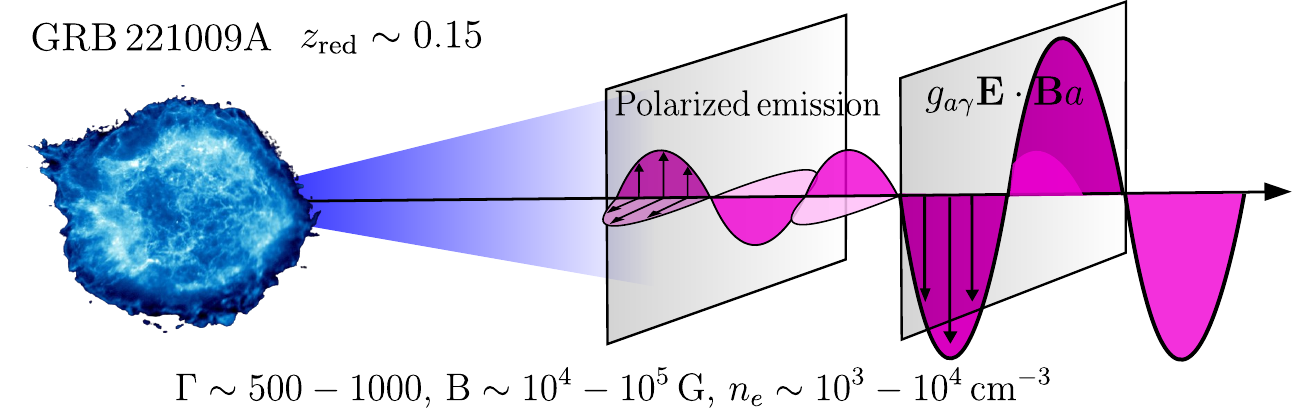}
	\caption{Schematic about gamma-ray burst polarized electromagnetic emission and its subsequent depolarization due to the axion field $a$ mixing with the electromagnetic field with strength $g_{a \gamma}$. The axion field may also alter the frequency and amplitude of the electromagnetic emission. Representative ranges of the astrophysical parameters from GRB 221009A are shown for reference.}
	\label{fig:GRB-diagram}
\end{figure}

Axion-photon conversions also induce a change in the polarization degree of the electromagnetic emission. A schematic representation of this effect in the context of gamma-ray burst (GRB) emission is shown in Fig.~\ref{fig:GRB-diagram}. This effect has been discussed in various works \cite{Rubbia:2007hf,Bassan:2010ya, Mena_2011,Gan:2023swl}, however, the impact of the axion field on the polarization of the emitted light has not been quantified using Stokes parameters, as it is commonly done in the astrophysics community \cite{2012ApJ...758L...1Y}.
In order to assess the impact of the axion on the polarization of the emitted radiation and properly compare it with current measurements from sources, a detailed calculation of the Stokes parameters is required.

GRBs are the most radiant explosions in the Universe and are considered to be highly captivating astrophysical events. These represent great potential as sources for multimessenger detection of nonelectromagnetic signals, including very high-energy (VHE) neutrinos, cosmic rays, and gravitational waves \citep[for a review, see][]{2015PhR...561....1K, 1999PhR...314..575P}.
Within an interval ranging from milliseconds to a few hours, events of this nature have the capability to discharge an enormous amount of energy, reaching up to $10^{55}\,{\rm erg}$ in the keV-MeV energy range  \citep{1993ApJ...413L.101K, 1993ApJ...413..281B}. Typically, this emission is mainly described by the empirical Band function \citep{1993ApJ...413..281B} or synchrotron radiation \citep{1998MNRAS.296..275D, 2011A&A...526A.110D}, which is expected from shock-accelerated electrons submerged in a magnetic field.   Although the origin of magnetic fields remains a subject of debate \citep[e.g., see][]{PhysRevLett.2.83,1980MNRAS.193..439L,2003astro.ph.12347L,2011ApJ...726...90Z,2011ApJ...726...75S}, one method to investigate their structure and origin is by analyzing the polarization properties. For the purpose of gathering data associated with GRB sources, a number of authors have employed polarization scenarios 
\citep[e.g., see][]{2003ApJ...594L..83G, Nakar, 2004MNRAS.354...86R, Gill-1,  2020ApJ...892..131S, 2021MNRAS.507.5340T, 2022MNRAS.tmp.2180S} to discuss the polarization degree found in GRB 090102 \cite{Steele}, GRB 120308A \cite{Mundell}, GRB 171205A \cite{2019ApJ...884L..58U},  GRB 190114C \cite{Laskar_2019}, GRB
190829A \cite{2022MNRAS.512.2337D}, and  GRB 191221B \cite{Buckley} among others.

Detecting TeV photons from bursts would offer vital insights into the GRB physics, particularly the potential involvement of hadronic, or leptonic processes \cite{2019ApJ...885...29F, 2019ApJ...883..162F}. It is thought that the nearest and most intense bursts would release photons with energies above TeV  \citep{2019ApJ...884..117W, 2019ApJ...883..162F}, although due to the attenuation with low-energy photons from the extragalactic background light \citep[EBL;][]{1966PhRvL..16..252G}, these photons can hardly reach the Earth. 

The Gamma-Ray Burst Monitor (GBM) instrument on board the {\itshape Fermi} satellite detected the extremely bright burst GRB 221009A at 13:16:59 UT on October 9, 2022 \citep{2022GCN.32636....1V, Lesage+23_221009a}. Hundreds of seconds later, a variety of instruments also detected the immediate release of radiation called ``prompt episode" \citep{gcn32637,gcn32641,2023ApJ...949L...7F,GRBAplha_221009A,Williams+23Swift_boat}.  It involved, for the first time,  the detection of VHE photons above 10 TeV by the Large High Altitude Air Shower Observatory \citep[LHAASO;][]{Huang+22LHASO_gcn32677, Cao+23LHAASO}.  GRB 221009A had several bright episodes that are detailed in \cite{Lesage+23_221009a}. In particular, the episodes associated with the intervals between $219$ and $278\,{\rm  s}$,  and  $517$ and $520\, {\rm s}$ after the trigger time, provided more than 90\% of the entire flux. The fluence of this extremely bright burst with $\sim 10^{-1}\, {\rm erg\, cm^{-2}}$ in the 10-1000 keV range, surpassed by more than twice the total amount of gamma-ray fluences from all bursts detected by Fermi-GBM; \citep[for details, see][]{Burns+23boat}.    Based on the absorption characteristic lines of CaII, CaI, and NaID, GRB 221009A was located at a redshift of $z_{\rm red} = 0.151$, with its corresponding luminosity distance of $D_L\approx 2.2\times10^{27}\, {\rm cm}$ \citep{deugarte+2222109a_redshift_gcn32648}. 

This GRB attracted particular attention of the astrophysics and astroparticle physics communities, due to the unexpected detection from this source of energetic photons higher than 10 TeV, which are above the maximum photon energy able to surpass the EBL, and some beyond the Standard Model explanations involving axion-photon conversion were proposed \cite{Troitsky:2023uwu, Wang:2023okw, Nakagawa:2022wwm}.
Here we focus solely on the polarization measurement of this GRB performed by the Imaging X-ray Polarimetry Explorer (IXPE) in the soft x-ray energy band, and use this information for the first time to set bounds on the photon-ALP coupling. For this purpose, we will use the polarization degree of GRB 221009A from one of the two-rings of prompt emission \cite{Negro:2023cer}, which reads $\Pi_{2-8\,\rm keV}=0.172\pm0.088$.
 
The paper is organized as follows: In Sec.~\ref{sec:polarization} we describe both the standard polarization formalism in the GRB scenario and the changes due to the addition of a photon-ALP coupling. In Sec.~\ref{sec:bounds} we present the bounds on the photon-ALP coupling and discussion of our results. Finally, in Sec.~\ref{sec:conclusions}, we provide the relevant conclusions. Henceforth, we will work with natural units $c=\hbar=k_B=1$ and use the cosmological parameters $\Omega_\Lambda=1-\Omega_m=0.685$, $H=67.4\, {\rm km~s^{-1} Mpc^{-1}}$ \citep{2020A&A...641A...6P}. In the context of our discussion, we write unprimed and primed terms to denote quantities in the observer and comoving frames, respectively.
\vspace{-0.8cm}
\section{GRB POLARIZATION}\label{sec:polarization}
\subsection{Standard scenario}\label{sec:std}
The expected polarization degree from a GRB can be obtained from its synchroton emission. The synchrotron power emitted per unit frequency from a single electron with Lorentz factor $\gamma_e^\prime$ gyrating in a magnetic field $\textbf{B}^\prime$ can be separated into a parallel and a perpendicular component as \cite{1959ApJ...130..241W,1986rpa..book.....R}
\begin{align}\label{eq:power_components}
    \begin{split}
        P_{\parallel}^\prime(\nu^\prime,\gamma_e^\prime)&=\frac{e^3{\nu^\prime}^2 B^\prime \sin\theta_p^\prime}{\gamma_e^\prime m_e}\int{\|A_{\parallel}(\nu^\prime,{\gamma_e^\prime},{\theta_e^\prime})\|^2d\theta_e^\prime},\cr
        P_{\perp}^\prime(\nu^\prime,\gamma_e^\prime)&=\frac{e^3{\nu^\prime}^2 B^\prime \sin\theta_p^\prime}{\gamma_e^\prime m_e}\int{\|A_{\perp}(\nu^\prime,{\gamma_e^\prime},{\theta_e^\prime})\|^2d\theta_e^\prime},
    \end{split}
\end{align}
where $A_{\perp}$ and $A_{\parallel}$ are the photon polarization components perpendicular and parallel to the projection of $\textbf{B}$ in the $x-y$ plane, respectively.
Here, primed symbols denote quantities in the GRB jet comoving frame, $m_e$ is the electron mass, $e$ is the elementary charge and $\theta_p^\prime$ is the pitch angle between the electron's velocity and the magnetic field.

The angle $\theta_e^\prime$ is the angle between the line of sight and the plane containing the electron trajectory.
The integration is performed from $-\pi$ to $\pi$. However, the amplitude squared of the polarization components is negligible outside $|\theta_e^\prime|\sim1/\gamma_e^\prime$, so the integral may be extended to the region $(-\infty,\infty)$, in which case analytic solutions can be found as 
\begin{align}\label{eq:power_components_analytic}
    \begin{split}
        P_{\parallel}^\prime(\nu^\prime,\gamma_e^\prime)=\frac{\sqrt{3}e^3B^\prime \sin\theta_p^\prime}{4\pi m_e}[F(x^\prime)-G(x^\prime)],\cr
        P_{\perp}^\prime(\nu^\prime,\gamma_e^\prime)=\frac{\sqrt{3}e^3B^\prime \sin\theta_p^\prime}{4\pi m_e}[F(x^\prime)+G(x^\prime)],
    \end{split}
\end{align}
where $x^\prime=\nu^\prime/\nu_c^\prime$ with the characteristic synchrotron frequency $\nu_c^\prime=3e B^\prime \sin\theta_p^\prime {\gamma_e^\prime}^2/(4\pi m_e)$ and
\begin{align}
    \begin{split}
        F(x)&=x\int_{x}^{\infty}{K_{\frac53}(u)du},\cr
        G(x)&=xK_{\frac23}(x),
    \end{split}
\end{align}
are in terms of the modified Bessel functions of the second kind $K_v(z)$. The total emitted power per unit frequency $P^\prime(\nu^\prime,\gamma_e^\prime)=P_{\parallel}^\prime(\nu^\prime,\gamma_e^\prime)+P_{\perp}^\prime(\nu^\prime,\gamma_e^\prime)$ is then readily found as the addition of the parallel and perpendicular components.

The synchrotron linear polarization degree for particles of a single energy $\gamma_e^\prime$ can then be readily calculated as \citep{1986rpa..book.....R}
\begin{equation}\label{eq:lin_def_pol_mono}
    \Pi_{\rm syn}^\prime(\nu^\prime,\gamma_e^\prime)=\frac{P_{\perp}^\prime(\nu^\prime,\gamma_e^\prime)-P_{\parallel}^\prime(\nu^\prime,\gamma_e^\prime)}{P_{\perp}^\prime(\nu^\prime,\gamma_e^\prime)+P_{\parallel}^\prime(\nu^\prime,\gamma_e^\prime)},
\end{equation}
where we take $\gamma_e^\prime$ to be the minimum Lorentz factor $\gamma_m^\prime$ of the electron energy distribution given by
\begin{equation}
    \gamma_m^\prime\approx4.2\times10^{4}\left(\frac{g(p)}{g(2.5)}\right)\left(\frac{\epsilon_e}{0.1}\right)\left(\frac{\Gamma}{700}\right),
\end{equation}
with $\Gamma$ the bulk Lorentz factor, $\epsilon_e$ the fraction of the emitted thermal energy that goes into the accelerated electrons, $g(p)=\left(\frac{p-2}{p-1}\right)$ and $p$ the spectral index of the electron distribution.

The synchrotron polarization for a given observed frequency $\nu$ can be derived by calculating the Stokes parameters \citep{2003ApJ...596L..17G,2009ApJ...698.1042T,2020MNRAS.498.3492C}
\begin{align}
\begin{Bmatrix}Q_\nu\\U_{\nu}\end{Bmatrix}=\frac{1+z_{\rm red}}{4\pi D_L^2}\int d\phi\int{d\cos\theta\, \delta_D^\varepsilon P^\prime(\nu^\prime,\gamma_e^\prime)\Pi_{\rm syn}^\prime\begin{Bmatrix}\cos2\chi\\ \sin2\chi\end{Bmatrix}},
\end{align}
and the spectral fluence \citep{2009ApJ...698.1042T}
\begin{equation}
    I_\nu=\frac{1+z_{\rm red}}{4\pi D_L^2}\int d\phi\int{d\cos\theta\, \delta_D^\varepsilon P^\prime(\nu^\prime,\gamma_e^\prime)}.
\end{equation}
Here, angular quantities are measured in the spherical coordinate system $(r,\theta,\phi)$ in the lab frame, where $\theta=0$ is the line of sight. The variable $\chi$ is the polarization position angle in the observer frame (see Refs. \cite{2009ApJ...698.1042T,2020MNRAS.498.3492C} for the explicit expression), $\delta_D=\frac{1}{\Gamma(1-\beta\cos\theta)}$ is the Doppler factor, $\nu^\prime=\frac{1+z_{\rm red}}{\delta_D}\nu$ is the comoving frequency, $z_{\rm red}$ is the redshift, $D_L$ is the luminosity distance to the source and $\varepsilon\in\{2,3\}$ is an index that represents instantaneous emission ($\varepsilon=3$, relevant for the afterglow) or time-integrated emission ($\varepsilon=2$, relevant for the prompt emission when integrated over a time larger than the duration of a single pulse and the value we use in the following) \citep{2003ApJ...596L..17G}.

We assume a GRB jet with half opening angle $\theta_j$ with bulk Lorentz factor $\Gamma$ and a thoroidal magnetic field. In this case, it is convenient to follow \cite{2009ApJ...698.1042T,2020MNRAS.498.3492C} and define the variables $y\equiv(\Gamma\theta)^2$, $y_j\equiv(\Gamma\theta_j)^2$ and $q\equiv\frac{\theta_v}{\theta_j}$, where $\theta_v$ is the viewing angle. We also consider small angles ($\cos\theta\approx1-\frac{1}{2}\theta^2$) and a relativistic jet ($1-\beta\approx\frac{1}{2\Gamma^2}$). We then have the Stokes parameters and spectral fluence
\begin{align}\label{eq:stokes_params}
    \begin{split}
        \begin{Bmatrix}Q_\nu\\U_{\nu}\end{Bmatrix}&=-\frac{1}{2}\frac{1+z_{\rm red}}{4\pi D_L^2}\frac{(2\Gamma)^\varepsilon}{\Gamma^2}\int_{0}^{y_j(1+q)^2}dy(1+y)^{-\varepsilon}\cr
        &\times\int_{-\Delta\phi(y)}^{\Delta\phi(y)}{d\phi P^\prime(\nu^\prime,\gamma_e^\prime,\phi)\Pi_{\rm syn}^\prime(\nu^\prime,\gamma_e^\prime,\phi)\begin{Bmatrix}\cos2\chi\\ \sin2\chi\end{Bmatrix}},\cr
        I_\nu&=-\frac{1}{2}\frac{1+z_{\rm red}}{4\pi D_L^2}\frac{(2\Gamma)^\varepsilon}{\Gamma^2}\int_{0}^{y_j(1+q)^2}dy(1+y)^{-\varepsilon}\cr
        &\times\int_{-\Delta\phi(y)}^{\Delta\phi(y)}{d\phi\,P^\prime(\nu^\prime,\gamma_e^\prime,\phi)},
    \end{split}
\end{align}
where $\Delta\phi(y)$ is provided in Eq.~(13) of \cite{2009ApJ...698.1042T}. The dependence of the emitted power per frequency $P^\prime(\nu^\prime,\gamma_e^\prime,\phi)$ on the angle $\phi$ is due to its dependence on the sine of pitch, which is given by \citep{2009ApJ...698.1042T,2020MNRAS.498.3492C}
\begin{widetext}
\vspace{0.3cm}
\begin{equation}
    \sin\theta_p^\prime(y,\phi)=\left[\left(\frac{1-y}{1+y}\right)^2+\frac{4y}{(1+y)^2}\frac{\left(\frac{\sqrt{y/y_j}}{q}-\cos\phi\right)^2}{\left(1+\frac{y/y_j}{q^2}-2\frac{\sqrt{y/y_j}}{q}\cos\phi\right)}\right]^\frac12 .
\end{equation}
\vspace{0.3cm}
\end{widetext}
Finally, for a given observation band $[\nu_1,\nu_2]$, the observed linear degree of polarization is
\begin{equation}\label{eq:lin_deg_pol_obs_band}
    \Pi_{\nu_1,\nu_2}(\gamma_e^\prime)=\frac{\sqrt{Q^2+U^2}}{I},
\end{equation}
where $S=\int_{\nu_1}^{\nu_2}{d\nu\, S_\nu}$ for $S\in\{Q,U,I\}$.
\vspace{-0.5cm}
\subsection{Polarization effects due to axion coupling to electromagnetic field in GRB}
\vspace{-.5cm}
The polarization components are modified by the addition of an interaction between the photon and the axion. The photon-ALP interaction Lagrangian reads \cite{PhysRevD.37.1237}
\begin{equation}
    \mathcal{L}_{a\gamma}=-\frac{1}{4}g_{a\gamma}F_{\mu\nu}\tilde{F}^{\mu\nu}a=g_{a\gamma}\textbf{E}\cdot\textbf{B},
\end{equation}
\vspace{0.1cm}
\noindent where $F_{\mu\nu}$ is the electromagnetic field tensor, $\tilde{F}_{\mu\nu}=\frac{1}{2}\epsilon_{\mu\nu\rho\sigma}F^{\rho\sigma}$ is its dual, $\textbf{E}$ and \textbf{B} are the electric and magnetic fields, respectively. The ALP field with a mass $m_a$ is denoted by $a$ and $g_{a\gamma}$ is the photon-ALP coupling.

The two photon polarization components are modified by the addition of this interaction as a function of the length of propagation from the GRB, which we assume propagates in the direction $z$. The evolution equations for a monoenergetic photon/ALP beam with energy $\omega$ are given by the system \cite{Bassan:2010ya,PhysRevD.37.1237,Mena_2011}
\begin{widetext}
\begin{align}\label{eq:system_alp}
\frac{d}{dz}\begin{pmatrix}A_{\perp,\rm ALP}(z)\\A_{\parallel,\rm ALP}(z)\\a(z)\end{pmatrix}=i\begin{pmatrix} \Delta_\perp\cos^2\xi+\Delta_\parallel\sin^2\xi & \cos\xi\sin\xi(\Delta_\parallel-\Delta_\perp) & \Delta_{a\gamma}\sin\xi\\ \cos\xi\sin\xi(\Delta_\parallel-\Delta_\perp) & \Delta_\perp\sin^2\xi+\Delta_\parallel\cos^2\xi & \Delta_{a\gamma}\cos\xi \\ \Delta_{a\gamma}\sin\xi & \Delta_{a\gamma}\cos\xi & \Delta_a \end{pmatrix}\begin{pmatrix}A_{\perp,\rm ALP}(z)\\A_{\parallel,\rm ALP}(z)\\a(z)\end{pmatrix}.
\end{align}
\end{widetext}
We assume that the magnetic field $\textbf{B}$ makes an angle $\xi\in[0,2\pi]$ with the $y$ axis in a fixed coordinate system. The polarization components $A_{\perp,\rm ALP}$ and $A_{\parallel,\rm ALP}$ are defined with respect to the magnetic field, where $A_{\perp,\rm ALP}$ is the component perpendicular to the direction of the magnetic field and $A_{\parallel,\rm ALP}$ is the component parallel to it. These components align with the $x$ and $y$ axes of the coordinate system when $\xi=0$, i. e. when the magnetic field aligns with the $y$-axis.

Following \cite{Bassan:2010ya}, the elements in the system of equations can be expressed as
\begin{align}\label{eq:defs_alp}
    \Delta_\perp\equiv2\Delta_{\rm QED}+\Delta_{\rm pl}&,\ \Delta_\parallel\equiv\frac72\Delta_{\rm QED}+\Delta_{\rm pl},\cr
    \Delta_{\mathrm{QED}}\equiv \frac{\alpha \omega^\prime}{45 \pi}\left(\frac{B^\prime}{B_{\mathrm{cr}}}\right)^2&,\ \Delta_{\mathrm{pl}}\equiv -\frac{\omega_{\mathrm{pl}}^2}{2 \omega^\prime},\cr
    \Delta_{a \gamma}\equiv \frac{1}{2} g_{a \gamma} B^\prime&,\ \Delta_a\equiv-\frac{m_a^2}{2 \omega^\prime},
\end{align}
where $\alpha$ is the fine structure constant and $B_{\rm cr}\equiv m_e^2/e$ is defined as the critical magnetic field. In this scenario, the photon beam propagates in a magnetized cold plasma,
which gives rise to an effective photon mass set by the plasma frequency $\omega_{\mathrm{pl}}=\sqrt{4 \pi \alpha n_e / m_e}$, where $n_e$ is the electron number density. 

The initial conditions of the system given by Eq.~(\ref{eq:system_alp}) are given by the polarization components of the no-ALP scenario
\begin{align}\label{eq:polarization_components}
    \begin{split}
        A_{\parallel,\mathrm{ALP}}^{z_0}=A_{\parallel}(\nu^\prime,{\gamma_e^\prime},{\theta_e^\prime})&=\frac{\sqrt{3}{\gamma_e^{\prime}}^2{\theta_e^\prime}}{2\pi\nu_c^\prime}\,v^\prime\,\theta_{\gamma}^\prime\ K_{\frac13} \left(\eta^\prime\right),\cr
        A_{\perp,\mathrm{ALP}}^{z_0}=A_{\perp}(\nu^\prime,{\gamma_e^\prime},{\theta_e^\prime})&=i\frac{\sqrt{3}\gamma_e^\prime}{2\pi\nu_c^\prime}\,v^\prime\,{\theta_\gamma^\prime}^2\ K_{\frac23} \left(\eta\prime\right),
        \end{split}
\end{align}
where $\theta_\gamma^\prime=\sqrt{1+{\gamma_e^\prime}^2{\theta_e^\prime}^2}$, $\eta^\prime=\frac{2\pi\nu^\prime a_c {\theta_{\gamma}^\prime}^3}{3{\gamma_e^\prime}^3}$, $a_c=\frac{3}{2}\frac{{\gamma_e^\prime}^3 v^\prime}{2\pi\nu_c^\prime}$ and $v^\prime=\sqrt{1-\frac{1}{{\gamma_e^\prime}^2}}$. In the case of the axion field, we have $a^{z_0}=0$. We solve these equations (with the definitions from Eq.~\ref{eq:defs_alp}) up to $z=R$ where $R\simeq2\Gamma^2 (1+z_{\rm red})^{-1}\delta t$ is the shock radius with variability timescale $\delta t$. Here, we have assumed that the photons are produced via the synchrotron process during the prompt episode of the burst~\cite{2019A&A...628A..59O,2011A&A...526A.110D,2013ApJ...769...69B}.  It has been shown that synchrotron emission could occur at in internal shocks at a significant distance from the source, typically in the range $10^{11}-10^{15}~\rm cm$~\cite{2005AIPC..784..164P,2014ApJ...794L...8B,2015MNRAS.450.2784F}. It is worth noting that some photospheric models have considered synchrotron emission at even smaller radii, around $10^9~\rm cm$. However, in the internal shock scenario, where multiple layers with different Lorentz factors collide, emission is expected to occur much farther away from the central engine~\cite{Sari:2000ks}. These collisions generate the observed variability in the GRB light curves and set the characteristic emission radius.

In our case, we assume that the synchrotron emission originates between $R_{\rm min}=z_0=10^{13}~\rm cm$ and $R$, which corresponds to a realistic internal shock model. This range accounts for interactions between shells with varying Lorentz factors, producing a small variability timescale consistent with observational constraints.

From here, we can obtain the power per unit frequency according to Eq.~(\ref{eq:power_components}) and follow the formalism detailed in Sec.~\ref{sec:std} to calculate the observed linear degree of polarization with effects from the photon-ALP interaction. In order to find the modified polarization angle $\chi$ due to the ALP-photon coupling, we diagonalize Eq.~\eqref{eq:system_alp}. We focus on the reduced system in the basis aligned with the external magnetic field $\mathbf{B}$, where only the perpendicular photon polarization mode $A_\perp$ mixes with the ALP field $a$:
\begin{equation}
\frac{d}{d z}\begin{pmatrix}A_{\perp,\rm ALP} \\a\end{pmatrix}=i\begin{pmatrix}\Delta_{\|} & \Delta_{a \gamma} \\\Delta_{a \gamma} & \Delta_a\end{pmatrix}\begin{pmatrix}A_{\perp,\rm ALP} \\a\end{pmatrix}\,.
\end{equation}
This system has eigenvalues
\begin{equation}
\lambda = \frac{\Delta_{\|}+\Delta_a}{2} \pm \frac{1}{2}\Delta_{\rm osc}\,,
\end{equation}
with oscillation wave number~\cite{Mena_2011,Bassan:2010ya,Huang:2022udc}
\begin{equation}\label{eq:Delta_osc}
\Delta_{\rm osc} \equiv \sqrt{(\Delta_a - \Delta_\parallel)^2 + 4\Delta_{a\gamma}^2}\,,
\end{equation}
and ALP-photon mixing angle
\begin{equation}
\tan(2\theta_{a\gamma}) = \frac{2 \Delta_{a \gamma}}{\Delta_{\|} - \Delta_a}\,.
\end{equation}
The perpendicular photon mode evolves as
\begin{align}
A_{\perp,\rm ALP}(z) &=A_{\perp,\mathrm{ALP}}^{z_0}\,e^{\frac{i(\Delta_{\|}+\Delta_a) z}{2}}\cr
&\hspace{-1.0cm}\times\left[\cos\left( \frac{\Delta_{\mathrm{osc}} z}{2} \right)+ i \cos(2\theta_{a\gamma}) \sin\left( \frac{\Delta_{\mathrm{osc}} z}{2} \right)\right]\,,
\end{align}
while the parallel mode evolves trivially as $A_{\parallel,\rm ALP}(z)=A_{\parallel,\mathrm{ALP}}^{z_0} e^{i \Delta_{\perp} z}$. We now consider a general initial polarization angle $\chi_0$ in the observer frame, and define $\xi$ as the projected angle of the transverse magnetic field on the plane of the sky. The polarization vector is then decomposed into components parallel and perpendicular to $\mathbf{B}$ as
\begin{equation}
\begin{pmatrix}A_{\parallel,\mathrm{ALP}}^{z_0} \\A_{\perp,\mathrm{ALP}}^{z_0}\end{pmatrix}\propto\begin{pmatrix}\cos(\chi_0 - \xi) \\\sin(\chi_0 - \xi)\end{pmatrix}\,.
\end{equation}
The modified polarization angle $\chi(z)$ is related to the relative phase between $A_{\parallel,\rm ALP}(z)$ and $A_{\perp,\rm ALP}(z)$, and is given by
\begin{equation}
\tan\left[2(\chi - \xi)\right]=\frac{2\,\mathrm{Re}(A_{\parallel,\rm ALP}^* A_{\perp,\rm ALP})}{|A_{\parallel,\rm ALP}|^2 - |A_{\perp,\rm ALP}|^2}\,.
\end{equation}
Substituting the expressions for $A_{\|}(z)$ and $A_{\perp}(z)$, we obtain the exact result:
\begin{equation}\small
\chi=\xi+\frac{1}{2}\arctan\left[\frac{\sin 2(\chi_0 - \xi)\,\cos(2\theta_{a\gamma})\,\tan\left(\frac{\Delta_{\rm osc} z}{2}\right)}{1 - \sin^2(\chi_0 - \xi)\,\left[1 - \cos(\Delta_{\rm osc} z)\right]}\right]\,.
\label{eq:chi_exact}
\end{equation}
This expression explicitly vanishes when the initial polarization is aligned or orthogonal to the magnetic field, i.e., $\chi_0 = \xi$ or $\chi_0 = \xi + \pi/2$. In the weak-mixing limit, $\theta_{a\gamma} \ll 1$, we can expand Eq.~\eqref{eq:chi_exact} to leading order in $\Delta_{a\gamma}/(\Delta_a - \Delta_\parallel)$, yielding
\begin{equation}\label{eq:chi_weak}
\chi=\chi_0+\frac{\Delta_{a\gamma}}{\Delta_a - \Delta_\parallel}
\sin(\Delta_{\rm osc} R)\sin\left[2(\chi_0 - \xi)\right]\,,
\end{equation}
where $\chi_0$ is the original polarization position angle in the observer frame in the standard astrophysical scenario (see Refs.~\cite{2009ApJ...698.1042T,2020MNRAS.498.3492C}).


\section{BOUNDS ON PHOTON-ALP COUPLING}\label{sec:bounds}
In this section we detail the procedure we performed to obtain bounds on the photon-ALP coupling via polarization measurements of GRB 221009A. 

We use the measurement of the linear polarization degree of GRB 221009A from one ring of prompt emission $\Pi_{2-8\,\rm keV}=0.172\pm0.088$ obtained with IXPE in the soft x-ray energy band ($2-8\,\mathrm{keV}$) \cite{Negro:2023cer}. The authors were also able to measure the x-ray polarization of the GRB afterglow, however the prompt emission has more relevance to us as the magnetic field strength of the afterglow is much weaker compared to the one of the prompt.

We can observe from Eqs.~(\ref{eq:stokes_params}) and (\ref{eq:lin_deg_pol_obs_band}) that the linear degree of polarization depends explicitly on the product of the jet opening angle and the bulk Lorentz factor $y_j$, and the ratio of the viewing angle to the jet opening angle $q=\theta_v/\theta_j$.
These parameters are, however, not determined which is why we have allowed some freedom in the choice of the bulk Lorentz factor and have worked with the benchmark values $\Gamma=500,700$ and $1000$. These are consistent with the afterglow modeling for the observed GeV, TeV and neutrino emission and the constraints derived on the minimum Lorentz factor \cite{Gao:2023csq,2023ApJ...943L...2L}. In the case of the jet opening angle, we take the base value of $\theta_j=1.5^\circ$ that gives a polarization consistent with the IXPE spectropolarimetric measurement \cite{Negro:2023cer}.
\begin{figure}[t!]
	\centering
	\includegraphics[width=0.95\linewidth]{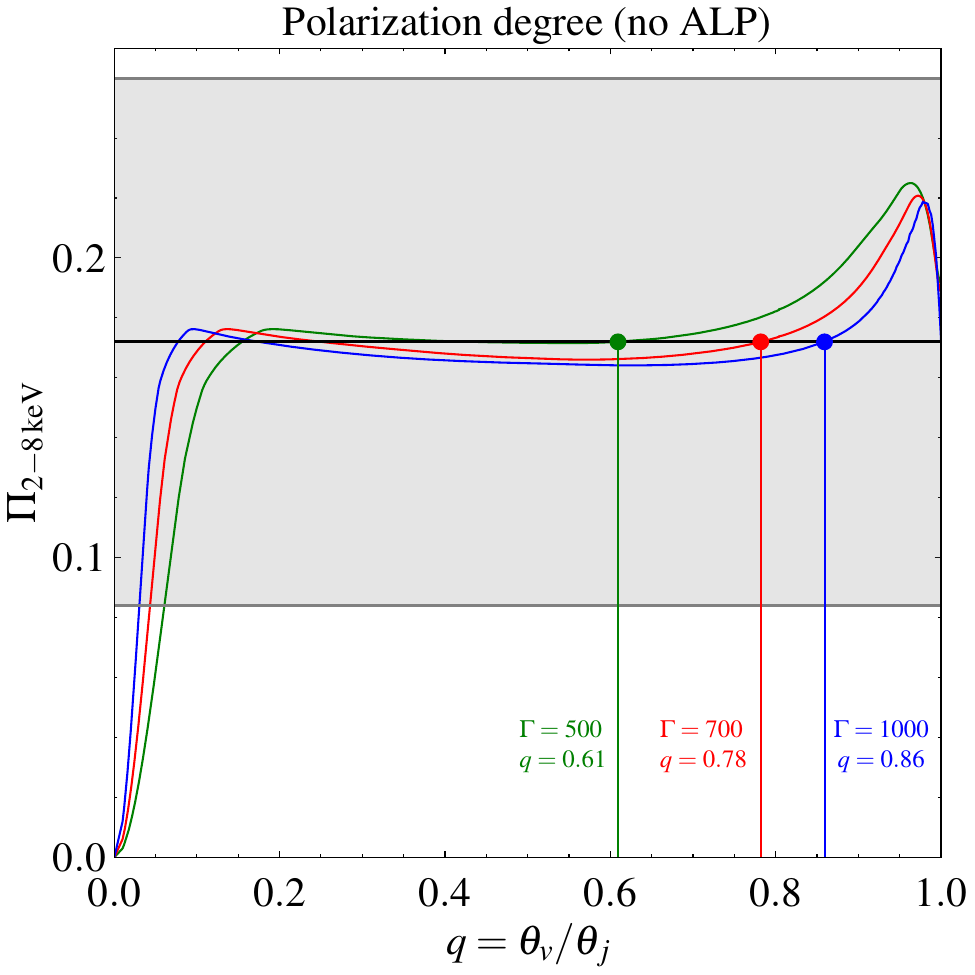}
	\caption{Linear polarization degree in the $2-8\,\rm keV$ band as a function of $q=\theta_v/\theta_j$, where $\theta_v$ is the observer's viewing angle and $\theta_j$ is the jet opening angle, for the three benchmark values of the bulk Lorentz factor $\Gamma=500,700,1000$. The horizontal black line and the gray region correspond to the IXPE measurement $\Pi_{2-8\,\rm keV}=0.172\pm0.088$. The colored dots show the value of $q$ for which the measurement can be explained by the GRB polarization model with no axion.}
	\label{fig:NoAxionPol}
\end{figure}

In order to fix $\theta_v$, or alternatively $q$, we calculate the linear polarization degree with no axion contribution according to Eq.~(\ref{eq:lin_deg_pol_obs_band}) as a function of $q$ and pick the value that reproduces the IXPE measurement. This procedure is exemplified in Fig.~\ref{fig:NoAxionPol} for the three benchmark values of the bulk Lorentz factor. As we can observe, there are two or three possible solutions for $q\in[0,1]$ depending on $\Gamma$, so we choose the one closest to $q=2/3$, which corresponds to the most likely configuration \cite{Ghisellini:1999xw}.
\begin{table}[h!]
\centering
\caption{Parameter values for the benchmark cases.}
\label{tab:params}
\begin{tabular}{|c|ccc|}
\hline
 & \multicolumn{1}{c|}{Benchmark 1} & \multicolumn{1}{c|}{Benchmark 2} & Benchmark 3 \\ \hline
 & \multicolumn{3}{c|}{$z_{\rm red}=0.151$, $\theta_j=1.5^\circ$} \\ \hline
$\Gamma$ & \multicolumn{1}{c|}{$500$} & \multicolumn{1}{c|}{$700$} & $1000$ \\ \hline
$y_j$ & \multicolumn{1}{c|}{$171.35$} & \multicolumn{1}{c|}{$335.84$} & $685.39$ \\ \hline
$q=\theta_v/\theta_j$ & \multicolumn{1}{c|}{$0.61$} & \multicolumn{1}{c|}{$0.78$} & $0.86$ \\ \hline
$B^\prime~[\mathrm{G}]$ & \multicolumn{1}{c|}{$1.4\times10^5$} & \multicolumn{1}{c|}{$10^5$} & $7\times10^4$ \\ \hline
$n_e~[\mathrm{cm}^{-3}]$ & \multicolumn{1}{c|}{$6.0\times10^4$} & \multicolumn{1}{c|}{$1.6\times10^4$} & $3.7\times10^3$ \\ \hline
$R~[\mathrm{cm}]$ & \multicolumn{1}{c|}{$1.23\times10^{15}$} & \multicolumn{1}{c|}{$2.41\times10^{15}$} & $4.92\times10^{15}$ \\ \hline
$\omega_{\mathrm{pl}}~[\mathrm{eV}]$ & \multicolumn{1}{c|}{$9.08\times10^{-9}$} & \multicolumn{1}{c|}{$4.63\times10^{-9}$} & $2.27\times10^{-9}$ \\ \hline
\end{tabular}
\end{table}

To obtain the aforementioned curves, we must also account for the electron number density surrounding the shock front. We assume that the forward shock propagates in a wind medium with number density $n_e(R)=A R^{-2}$ with $A=3.02\times10^{35}A_\star~\mathrm{cm}^{-1}$, where we set $A_\star=0.3$ according to the estimate performed by \cite{Negro:2023cer}. As mentioned previously, the shock radius is calculated as $R=2\Gamma^2\delta t$, where we take the variability timescale value to be $\delta t=0.082\ \rm s$ \citep{2023SCPMA..6689511W}. Finally, we calculate the magnetic field strength as $B^\prime\sim10^{5}\left(\frac{\Gamma}{700}\right)^{-1}$.
We summarize the values of all the parameters for the benchmark cases in Table \ref{tab:params}.

\begin{figure}[h]
	\centering
\includegraphics[width=0.95\linewidth]{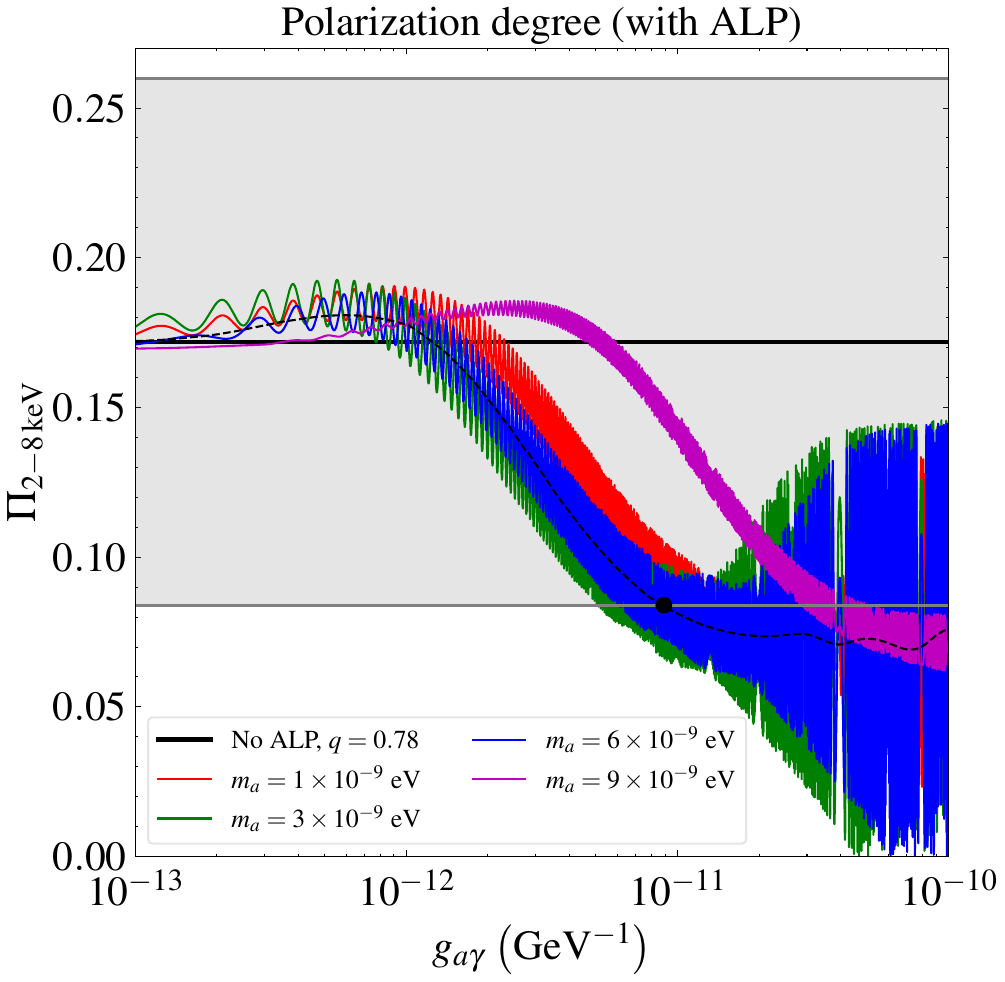}
	\caption{Polarization degree of GRB 21009A as a function of the ALP-photon coupling strength, for four values of the axion mass and parameters according to benchmark 2. For comparison, we show the measured value of the polarization degree with a horizontal black line and uncertainty represented by the gray region. The dashed black curve corresponds to the application of a one-dimensional Gaussian filter on the blue solution and the black dot represents the point when the dashed curve crosses past the uncertainty. It can be clearly appreciated that at sufficiently large couplings, the electromagnetic emission can be depolarized below current experimental uncertainty.}
	\label{fig:PiVsg_agamma}
\end{figure}

Once all parameters for each of the benchmark cases have been fixed, we are able to test how the linear degree of polarization is modified by the addition of the ALP. We show this effect in Fig.~\ref{fig:PiVsg_agamma}, where we present the polarization degree measured from GRB 221009A with a horizontal black line and the uncertainty with the corresponding gray region. As explained previously, the expected polarization degree induced by synchroton emission lies well within the experimental uncertainties and can be explained with the standard, no-ALP model.

In addition, we show how the polarization degree induced by an axion field with masses $m_a=10^{-9}$ eV (red), $m_a= 3 \times 10^{-9}$ eV (green), $m_a= 6\times10^{-9}$ eV (blue) and $m_a= 9 \times 10^{-9}$ eV (purple) evolves for different values of the photon-ALP coupling, and GRB parameters according to benchmark 2. We fix the angle $\xi=\pi/2$, which corresponds to the magnetic field aligned in such a way that it results in maximal parallel polarization, corresponding to the highest degree of polarization. It can be appreciated in the figure that, for an axion mass of $m_a= 6 \times 10^{-9}$ eV,  values of the axion-photon coupling above $g_{a \gamma} \gtrsim 7 \times 10^{-12}$ GeV$^{-1}$ would depolarize the electromagnetic emission below the experimental uncertainty.\\

The oscillations in the polarization degree stem from the oscillating nature of the axion field itself and more precisely from the form of the probability that a polarized photon converts into an ALP after a distance $d$, which scales as a quadratic sinusoidal function whose amplitude, minima and maxima depend on the value of the photon-ALP coupling \cite{Mena_2011,Bassan:2010ya,Huang:2022udc}
\begin{equation}
    P_{a\gamma}=\sin^2(2\theta_{a\gamma})\sin^2\left(\frac{\Delta_{\rm osc}d}{2}\right).
\end{equation}
Further analysis of the conversion probability allows one to define the critical energies \cite{DeAngelis:2007wiw,Bassan:2010ya}
\begin{align}\label{eq:crit_energies}
    \begin{split}
        \omega_L^\prime&\equiv\frac{\omega^\prime|\Delta_a-\Delta_{\rm pl}|}{2\Delta_{a\gamma}}\cr
        &\approx2.5\times10^{-2}~\mathrm{eV}\left[\Bigg|\left(\frac{m_a}{10^{-9}~\rm eV}\right)^2-\left(\frac{\omega_{\rm pl}}{10^{-9}~\rm eV}\right)^2\Bigg|\right]\cr
        &\hspace{0.5cm}\times\left(\frac{10^5~\rm G}{B^\prime}\right)\left(\frac{10^{-11}~\mathrm{GeV}^{-1}}{g_{a\gamma}}\right),\cr
        \omega_H^\prime&=\frac{90\pi g_{a\gamma}B_{\rm cr}^2}{7\alpha B^\prime}\cr
        &\approx21~\mathrm{keV}\left(\frac{10^5~\rm G}{B^\prime}\right)\left(\frac{g_{a\gamma}}{10^{-11}~\mathrm{GeV}^{-1}}\right).
    \end{split}
\end{align}
In the energy range $\omega_L^\prime<\omega^\prime<\omega_H^\prime$ the photon-ALP mixing is maximal ($\theta_{a\gamma}\approx\pi/4$) and the conversion probability becomes energy-independent. This is the so-called \textit{strong-mixing regime}. Outside of this energy range, the conversion probability becomes energy-dependent and very small, so that $\omega_L^\prime$ and $\omega_H^\prime$ can be understood as the low- and high-energy energy cutoff, respectively.

These two critical energies are fundamental to understand the behavior of the polarization degree shown in Fig.~\ref{fig:PiVsg_agamma}. First, we recall that this quantity is calculated according to Eqs.~(\ref{eq:stokes_params}) and (\ref{eq:lin_deg_pol_obs_band}), as integrals of the Stokes parameters and spectral fluence over the observed energies; in the case of the IXPE observations $\omega_1=2~\rm keV$ and $\omega_2=8~\rm keV$. However, the Stokes parameters depend on $P^\prime$ and $\Pi_{\rm syn}^\prime$ which themselves are functions of the energy measured in the GRB jet comoving frame, which transforms according to
\begin{align}\label{eq:energy_transform}
    \omega_i^\prime&\approx\frac{(1+z_{\rm red})(1+y)}{2\Gamma}\cr
    &=0.49~\mathrm{keV}\left(\frac{1+z_{\rm red}}{1+0.151}\right)\left(\frac{1+y}{1+300}\right)\cr
    &\hspace{1.7cm}\times\left(\frac{700}{\Gamma}\right)\left(\frac{\omega_i}{2~\rm keV}\right).
\end{align}
For a fixed ALP mass, we notice from Eq.~(\ref{eq:crit_energies}) that $\omega_L^\prime$ decreases and $\omega_H^\prime$ increases as the coupling $g_{a\gamma}$ becomes larger. This effectively means the strong-mixing regime broadens with the strength of the interaction, in line with the physical expectation. The broader this regime, the broader is the region of $y\equiv\Gamma\theta$ values satisfying $\omega_L^\prime<\omega_1^\prime<\omega_2^\prime<\omega_H^\prime$, so the contribution of the ALP effect increases in the Stokes parameters $dy$ integral. This explains why as $g_{a\gamma}$ grows, the oscillations and amplitude in each of the colored curves of Fig.~\ref{fig:PiVsg_agamma} do as well. Following the same argument, as the coupling drops, the frequency of the oscillations and the amplitude do too and we approach the no-ALP scenario for $g_{a\gamma}\lesssim10^{-13}~\mathrm{GeV}^{-1}$.

We further notice that if the ALP mass is equal to the plasma frequency ($m_a=\omega_{\rm pl}$), then by Eq.~(\ref{eq:crit_energies}) $\omega_L^\prime=0$ and the strong-mixing regime reaches its maximum width. This corresponds to a resonance; the mass for which the ALP effect on the polarization is maximal. We can also appreciate this in Fig.~\ref{fig:PiVsg_agamma}. We recall that the plasma frequency for the parameters used in this figure is $\omega_{\rm pl}\sim4.6\times10^{-9}~\rm eV$. For $m_a<\omega_{\rm pl}$ we have the blue curve, whose oscillation amplitude and frequency are smaller than those of the green curve, which satisfies $m_a>\omega_{\rm pl}$ and also has a mass closer to $\omega_{\rm pl}$. 

We would also like to draw attention to the purple curve, which has the smallest oscillation amplitude compared to all the others. This is because this solution has the largest ALP mass, which increases the value of $\omega_L^\prime$ and decreases the width of the strong-mixing regime. This effectively means that the ALP contribution decreases, as we have discussed previously.

We will now describe the criteria for deriving a bound on the ALP-photon coupling for a given ALP mass. The observed GRB light curves constrain the possible range of bulk Lorentz factors~\cite{Derishev:2023xyx,OConnor:2023ieu}, which agree with our benchmark values from table~\ref{tab:params}. Using this information, we compute the expected polarization degree in the standard (no-ALP) scenario and find that the viewing angle ratio that best fits the observations is $q\sim2/3$, consistent with the theoretical prediction from standard astrophysical models. Given these constraints, it is reasonable to assume that the initial polarization degree in the absence of ALPs corresponds to the measurement from IXPE, as the standard model expectation should match the observed polarization. At very small values of the ALP-photon coupling $g_{a\gamma}$, the polarization degree is expected to be that of the no-ALP scenario, as the ALP-induced effects become negligible. As the coupling strength increases, the polarization degree deviates more significantly from the no-ALP scenario, and we set a bound only when the polarization degree first exceeds the experimental uncertainty.

However, due to the oscillating nature of the ALP-photon system, the polarization degree exhibits oscillations, as shown in Fig.~\ref{fig:PiVsg_agamma}. For large enough ALP-photon coupling strengths (e.g., $g_{a\gamma}\gtrsim4\times10^{-12}~\mathrm{GeV}^{-1}$), the oscillation length $\ell_{\rm osc}=2\pi/\Delta_{\rm osc}$ (see Eq.~\eqref{eq:Delta_osc}) becomes much smaller than the propagation distance $R$, resulting in a high-frequency oscillation in the polarization degree. In this regime, small changes in the coupling strength $g_{a\gamma}$ lead to significant differences in the polarization degree, with one value potentially exceeding the experimental uncertainty while a nearby value does not.

To address this sensitivity and the rapid oscillations, we apply a one-dimensional Gaussian filter to average the polarization degree over the oscillation length. This approach reflects the fact that the resolution of the measurement is not fine enough to resolve such rapid oscillations, effectively smoothing out these variations, as would be observed in the data. The dashed black curve in the figure illustrates the result of this averaging procedure, which allows us to set not so restrictive constraints on the coupling strength despite the oscillatory behavior.

Finally, we would also like to mention that, in order to take into account the uncertainty on the angle $\xi$, we perform the procedure detailed above for a random sample of $\xi$. Then, we set the bounds as an average over this sample.

\begin{figure}[t!]
	\centering
	\includegraphics[width=0.95\linewidth]{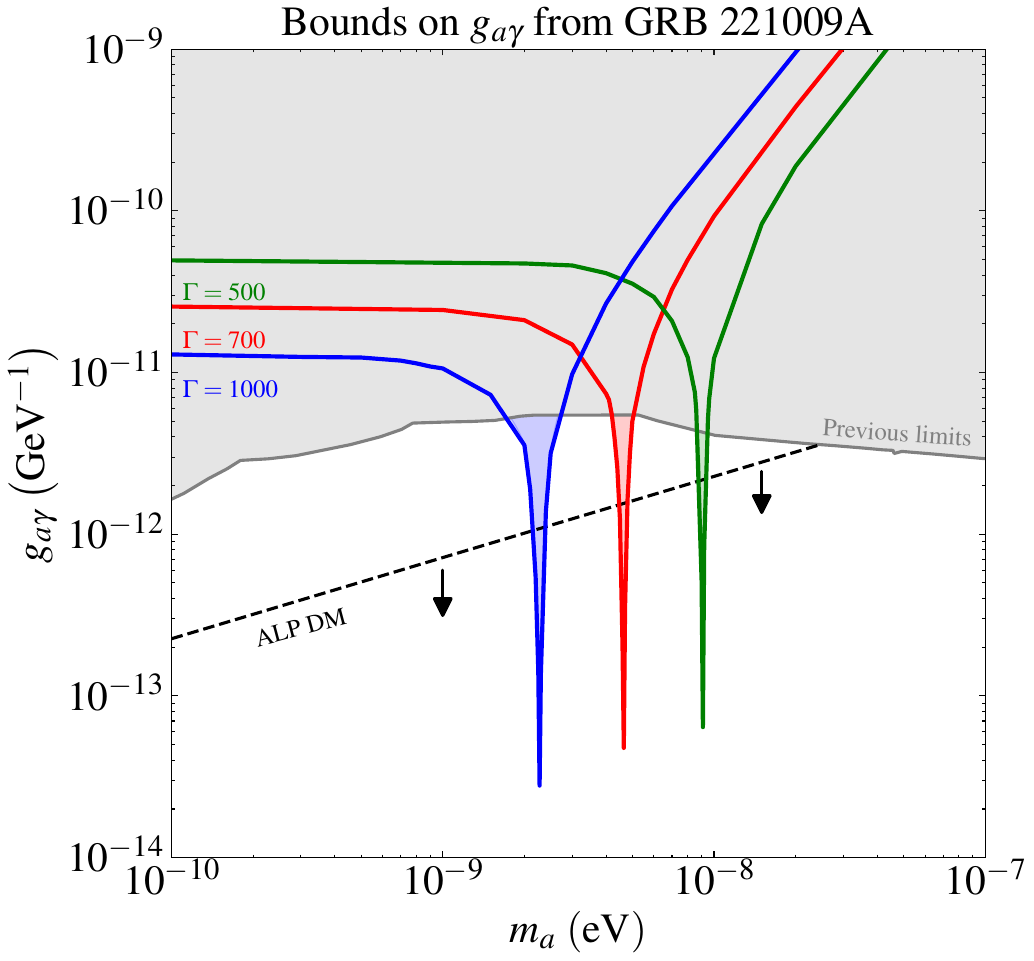}
	\caption{Bounds on ALP-photon coupling as a function of the ALP mass from GRB 221009A for three benchmark cases. The gray region shows previous bounds from a variety of complementary searches \cite{Li:2020pcn,Pant:2023lnz,Ning:2024eky,2024arXiv240519393M,Dessert:2022yqq,Noordhuis:2022ljw, Escudero:2023vgv, Eckner:2022rwf}. We also plot the region favored by ALP dark matter below the dashed black line \cite{Essig:2013lka}.}
	\label{fig:axion_constraint}
\end{figure}

In Fig.~\ref{fig:axion_constraint}, we present the result of applying our criteria to derive bounds in the $g_{a\gamma}-m_a$ plane on the ALP-photon coupling for the three benchmark cases we consider. For comparison, we show the combination of bounds from a variety of complementary searches: stellar axion production in the M82 starburst galaxy \cite{Ning:2024eky}, invisible neutrino decay into ALPs from NGC 1068 \cite{Pant:2023lnz}, ALP-photon oscillation effect in the spectra of the blazar Markarian 421 (Mrk 421) \cite{Li:2020pcn}, conversion of extragalactic gamma-rays into ALPs and subsequent reconversion in the Milky Way's magnetic field with sub-PeV gamma-ray observations \cite{Eckner:2022rwf}, axion-photon conversion on the magnetic field of the progenitor star of SN1987A \cite{2024arXiv240519393M}, magnetic white dwarf polarization measurements \cite{Dessert:2022yqq}, axion production in neutron star magnetospheres \cite{Noordhuis:2022ljw} and the impact of axion star decays on the reionization of the intergalactic medium during the dark ages, comparing this effect with Planck measurements of the CMB optical depth \cite{Escudero:2023vgv}. We also present, below the dashed line, the region of parameter space that is favored by ALP dark matter and could account for all the DM produced either thermally in the big bang or nonthermally by the misalignment mechanism \cite{Essig:2013lka}.

There are several points to note in this figure. First, for small ALP masses we can see that the limits become constant. This is readily explained by the $\omega_L^\prime$ expression in Eq.~(\ref{eq:crit_energies}). For $m_a\ll\omega_{\rm pl}$, the dependence on the ALP mass of the critical frequency disappears and it is instead determined by the plasma frequency. This effectively means that the strong-mixing regime is independent of the ALP mass and so is the polarization degree, which translates into constant limits.

On the other hand, we note that the bounds are the strongest for $m_a=\omega_{\rm pl}$, which are represented by the minima of the curves at $g_{a\gamma}\sim10^{-14}~\mathrm{GeV}^{-1}$. As mentioned previously, this corresponds to the resonant case when the ALP effect on the polarization is maximal, so this behavior was to be expected. This also leads to the conclusion that our methodology will provide the best constraints around the region where $m_a\approx \omega_{\rm pl}$. This means that the determination of the circumburst electron number density of the GRB $n_e$ is incredibly important, as this determines the plasma frequency. While in this work, we have focused on the polarization from the prompt emission, it is possible to apply our formalism to afterglow polarization, where $n_e$ might be as small as $\sim10^{-6}~\mathrm{cm}^{-3}$ \cite{Fong:2015oha}, which would allow to set limits down to $m_a\sim10^{-14}~\rm eV$.

On the other hand, in the case when $m_a \gg\omega$, we see that the limits worsen as $m_a$ increases. The reason goes back once again to $\omega_L^\prime$. For large ALP masses, this lower critical frequency increases, which reduces the strong-mixing region quadratically, effectively making the ALP effect smaller. In fact, the limits shown in Fig.~\ref{fig:axion_constraint} also worsen quadratically in this region.

Regarding the differences between each benchmark, we notice that the bounds improve as the bulk Lorentz factor increases. We know that the magnetic field $B^\prime$ decreases as $\Gamma$ increases and according to Eq.~(\ref{eq:crit_energies}), as both critical frequencies are proportional to ${B^\prime}^{-1}$, this would imply a strong-mixing region of the same width, but shifted to higher energies. This is in line with the physical intuition that if the magnetic field becomes weaker, the beam energy must be larger in order to have strong conversion. 
Furthermore, from the energy transformation Eq.~(\ref{eq:energy_transform}) and the definition of $y\propto\Gamma^2$, we see that the dependence of the energy measured from the jet comoving frame scales as $\omega_i^\prime\propto\Gamma$, which means this one also is shifted to higher energies.
Then, the reason why the bounds improve as the bulk Lorentz factor increases is independent from the strong-mixing regime. Rather, it is simply from the Stokes parameters, where the integration region in $dy$ grows quadratically thanks to the dependence of the upper limit on $y_j\propto\Gamma^2$.

From Fig.~\ref{fig:axion_constraint} we can observe that our method allows us to probe and set competitive limits close to the resonance region $m_a\approx\omega_{\rm pl}$, reaching limits of the order $10^{-14}~\mathrm{GeV}^{-1}$ at resonance. However, we must also notice that the determination of GRB parameters such as the bulk Lorentz factor and the magnetic field strength are crucial for our method. By presenting our results for the three benchmark cases we consider, we highlight that uncertainty on any of the GRB parameters leads to uncertainty on the strength of our bounds and the location of the resonance, so in order to make any conclusions we require accurate information on the GRB.

\section{CONCLUSIONS}\label{sec:conclusions}

In this work we have improved the formalism for the calculation of the linear degree of polarization from GRBs due to photon-ALP interactions, and derived bounds from polarization measurements of GRB 221009A. For the first time, we have calculated the expected polarization induced by axion-photon coupling accounting for the Stokes parameters and spectral fluence, as it is typically done in the astrophysics literature.

We have shown how the polarization degree changes due to the introduction of this novel interaction and discussed its behavior based on the photon-ALP probability conversion and the strong-mixing regime. Further, we compared our expected polarization degree from synchroton radiation models including an ALP field with that observed from the IXPE instrument from GRB 221009A, accounting for the uncertainty in the measurement.

Based on IXPE observations of polarization from the prompt emission of GRB 221009A, we have introduced a method that allows us to set limits on the ALP-photon coupling. This method is strongest for ALP masses in the range $m_a\in[10^{-9},10^{-8}]~\rm eV$. However, we note that uncertainties in the GRB jet parameters such as the bulk Lorentz factor and magnetic field strength translate into uncertainties in our limits.
Our method is well suited for observation of small polarization degree ($\leq25\%$). In order to obtain the best possible constraints, the measurement should be as precise as possible, i.e., smallest uncertainty. To exemplify this point, we note that, if the IXPE observations, which have an uncertainty of $\sim50\%$, improved to $\sim25\%$, we would be able to set limits an order of magnitude stronger.

Considering a population of sources may also allow to enhance the sensitivity. In this work we have restricted ourselves to a single source as a proof of concept; an analysis including a population of sources will be completed in future work. Furthermore, such an analysis may be extended to other astrophysical objects from which polarization measurements have been obtained, such as active galactic nuclei, supernovae or neutron stars. The polarization measurements from these sources correspond to different wavelengths, thus allowing to probe different values of the ALP mass.
Future GRB polarization measurements will reduce the uncertainties in the observed polarization degree, and the bounds will become stronger correspondingly. Given the uncertainties on the emitted electromagnetic spectrum and initial polarization degrees, time-dependent information on the polarization degree may be particularly useful. An afterglow polarization consistent with synchroton radiation and no axion field, but inconsistent with the prompt emission polarization may be an indication of new physics.

\begin{section}{ACKNOWLEDGEMENTS} 
We are indebted to Merlin Reichard and Antonio Galván-Gámez for substantial numerical support. This research was performed using
services/resources provided by the LAMOD-UNAM project using the clusters Atocatl and Tochtli. LAMOD is a
collaborative effort between DGTIC and the research Institutes of Astronomy, Nuclear Science and chemistry at UNAM.  N.F. is grateful to UNAM-DGAPA-PAPIIT for the funding provided by Grant No. IN112525.  The work of G.H. is supported by the U.S. Department of Energy Office of Science under award number DE-SC0020262.

\end{section}

\bibliography{References}    

\end{document}